\documentclass{IEEEcsmag}

\usepackage[colorlinks,urlcolor=blue,linkcolor=blue,citecolor=blue]{hyperref}

\usepackage{upmath}

\begin{document}


\title{Extending the team with a project-specific bot}

\author{Théo Zimmermann}
\affil{Inria, Université de Paris, CNRS, IRIF, Paris, France}

\author{Julien Coolen}
\affil{Université de Paris, Paris, France}

\author{Jason Gross}
\affil{MIT, EECS department, Boston, MA, USA}

\author{Pierre-Marie Pédrot}
\affil{Inria and LS2N, Nantes, France}

\author{Gaëtan Gilbert}
\affil{Inria and LS2N, Nantes, France}


\begin{abstract}
While every other software team is adopting off-the-shelf bots to automate everyday tasks, the Coq team has made a different choice by developing and maintaining a project-specific bot from the ground up. In this article, we describe the reasons for this choice, what kind of automation this has allowed us to implement, how the many features of this custom bot have evolved based on internal feedback, and the technology and architecture choices that have made it possible.
\end{abstract}

\maketitle

\chapterinitial{On collaborative coding platforms} like GitHub, bots are commonplace nowadays~\cite{wessel2018power}, as it has become easy and encouraged to add new bots to one's projects.
This is great for small teams or single developers wanting to quickly speed up their project's adoption of best practices. But for larger teams with well-established processes, having to adapt to rigid workflows of preexisting bots can be disruptive.

Previous research has already shown that task-oriented GitHub bots can cause friction by lacking social context or disrupting developers' workflows~\cite{brown_sorry_2019}. Wessel and Steinmacher~\cite{wessel_inconvenient_2020} have proposed the promising concept of a meta-bot (aggregating and summarizing information coming from several bots) to alleviate these issues. For several years, we have explored another strategy, that shares some characteristics with a meta-bot: relying on a multi-task, project-specific bot, directly developed and maintained by the project team.
The bot works hand-in-hand with developers, helping them by automating everything that is repetitive and easily automatable, without requiring changes to their workflow. For medium to large teams, this can be a reasonable investment to make, that will be largely compensated by the returns. Besides, this does not preclude also using off-the-shelf solutions when they do match the team's needs.

We have adopted this strategy for the maintenance of the Coq proof assistant~\cite{the_coq_development_team_2021_5704840}, a medium-sized open-source software system, managed by a core team of about 10 developers, an extended maintainer team of about 30 people, and hundreds of contributors.
In this paper, we describe how we have developed and adopted such a bot, the tasks that it helps with, technical choices regarding its implementation, and how this helps maintain and evolve its code to assist the team.

\section{MAKING ROOM FOR A BOT}

We started the implementation of the Coq bot (\url{https://github.com/coq/bot}) in 2018. Although the trigger for writing this bot was the need for a specific feature, which did not exist in any preexisting solution at that time, it was conceived from the beginning as a multi-task bot that would evolve to assist the team. This is what has happened, leading to various new features over time.

\subsection{Bridging GitHub and GitLab for continuous testing}

The initial feature provided by this bot was a synchronization mechanism between pull requests [PRs] opened on a GitHub repository and branches on a GitLab mirror.

\subsubsection{Context}

In 2018, the Coq team was pushing the limits of Continuous Integration [CI] services to perform Reverse Dependency Compatibility Testing~\cite{zimmermann:tel-02451322,ochoa:hal-03420593}. This required advanced features like artifact sharing and job parallelization that few CI solutions provided at this time (this was before GitHub Actions~\cite{github_actions} were introduced). After trying out various providers, the team concluded that only GitLab CI was sufficiently advanced, with a sufficiently generous offer for open-source.
They had just introduced a feature to use their CI for GitHub repositories, but it was (and still is) limited because it does not synchronize PRs coming from GitHub forks.

\subsubsection{Feature}

The first task of this bot was thus to perform this synchronization: pushing and updating branches on the GitLab mirror for any opened or updated PR on the Coq GitHub repository.
Even though GitLab supports reporting CI results back to GitHub, the bot handles this as well. Initially, this allowed getting better names for CI statuses, and direct links to failed jobs in case of a pipeline failure. But the flexibility that this provided allowed several other improvements over the following months and years.

\subsubsection{Feedback and evolution}

Some Coq maintainers worried that CI could provide false confidence on PR impact after merge because the PR branch can be seriously lagging behind the base.
A first solution that was implemented in the bot to avoid this problem was to require the PR to be up-to-date with the base branch to run the CI at all. But after just a few weeks, feedback on this approach was very negative because it meant that developers had to rebase all the time, given the rapid integration of changes in the base.
A new solution was found and is still used to this day: the synchronization mechanism automatically creates merge commits between the PR head and the head of the base branch (similarly to Travis CI which also tests merge candidates).
Controlling the status report to GitHub was essential to implement this solution, since the bot can map from the tested merge commit to the origin GitHub commit.

When the Checks tab was introduced~\cite{github_checks_announcement}, we started relying on it to report CI log summaries directly on GitHub. Because the bot is project-specific, we can automatically search for errors in CI logs (based on knowledge of their expected shape) to ensure that we display them.

Following suggestions from reviewers, CI reports also include direct links to HTML documentation CI artifacts to ease previewing of documentation modifications.

While this bot is project-specific, and this feature is customized to be particularly suited for the Coq project, its core is of general interest and has been used beyond the repositories maintained by the Coq team.
Most of the other users are from the Coq ecosystem (e.g., the MathComp library), but some come from outside (e.g., the saltstack-formulas organization's hundreds of repositories).

\subsection{Keeping track of PRs with merge conflicts}

\subsubsection{Context}

Since the Coq team started using GitHub, developers gradually introduced labels until some point when they were reorganized into several categories, one of them being the \texttt{needs} labels, which indicate that something needs to happen before the PR can be merged.

\subsubsection{Feature}

When a new or updated PR cannot be merged cleanly with its base branch, CI cannot run, and instead, the bot adds a \texttt{needs: rebase} label. This information is quite useful to both authors and reviewers.

\subsubsection{Feedback and evolution}

The \texttt{needs} labels are very useful to maintainers skimming over the list of opened PRs, but sometimes PRs were missing the \texttt{needs: rebase} label because the merge conflict resulted from a change in the base branch after the last PR update.

To complete the information on PRs with conflicts, we introduced a preexisting GitHub Action that runs after each base branch update. In this case, a tool matching our use case was available, so there was no need to extend the bot.

Much more recently, the team decided to reduce the number of stale open PRs by having the bot automatically close PRs with the \texttt{needs: rebase} label set for more than 30 days, after a warning and an additional 30-day grace period. This is close to what ``stale bots'' implement~\cite{wessel_should_2019}, but with a different criterion to determine that a PR is stale, since it relies on merge conflicts rather than the absence of any activity. While it means that some PRs are not considered as stale even if they have been inactive for a while, this also means that the required action to remove the stale status is more demanding than just posting a comment (it requires solving merge conflicts).
Being in control of our (project-specific) bot was key to implementing this envisioned workflow.


\subsection{Merging pull requests}

\subsubsection{Context}

The Coq team has precise rules on when and how to merge a PR. We have already mentioned the absence of a \texttt{needs} label. There are additional requirements in terms of labels, milestones, assignees, reviews, target branches. Furthermore, the PR must be merged with a merge commit (instead of squashing, rebasing, or fast-forwarding), which must have a message in a specific format and be PGP-signed.
For several years, a merge script was available to check these requirements and apply the required formatting. However, it still represented a barrier to onboarding new maintainers (especially because of the requirement for signed merge commits).

\subsubsection{Feature}

We added support allowing maintainers to request the bot to merge a PR. The bot then checks that all requirements are met and that the maintainer is an authorized maintainer before performing the merge (see \textbf{Figure \ref{fig:merge-backport}}). Internally, it relies on GitHub's merge button to produce a signed merge commit, but it checks many things that this merge button alone would not check and uses the expected formatting.

\subsubsection{Feedback and evolution}

Since it was introduced, this has been the dominant method for merging PRs, by all maintainers. Some new maintainers have never called the merge script (that is still available as an alternative, for now).

The feature received only minor changes to provide complete feedback to maintainers who forget about several criteria when merging a PR (instead of failing for one reason, then failing for the next reason at the next attempt) and to take edited comments into account.

\subsection{Keeping track of the backporting process}

\begin{figure*}
\centerline{\includegraphics[width=34pc]{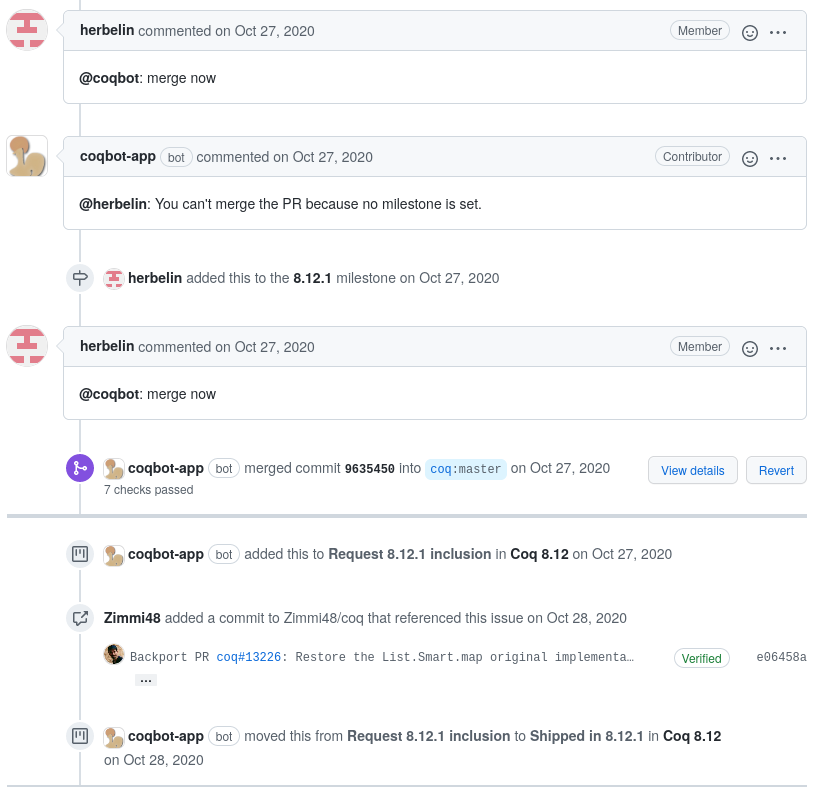}}
\caption{This screenshot demonstrates two features of the Coq bot. First, a maintainer uses the bot to merge a PR, but the bot reminds the maintainer that they have forgotten to set a milestone. After this is fixed, all the criteria for merging a PR are met, so the bot executes the command. Second, the bot analyzes the milestone of the merged PR to figure out if backporting was requested. In this case, it was, so it adds the PR to the appropriate ``Backport requested'' column of the RM backporting project. The RM then prepares the backport (on their fork), and when they push it to the release branch, the bot moves the PR to the corresponding ``Shipped'' column of the backporting project.}
\label{fig:merge-backport}
\end{figure*}

\subsubsection{Context}

Starting in 2017, in the context of a switch to shorter release cycles, the Coq team decided to change its branch management process to streamline it and facilitate contributions~\cite{zimmermann:tel-02451322}. Instead of pushing bug fixes to release branches and merging these branches regularly into the development branch, it was decided that PRs should always target the development branch, then be backported to release branches (if relevant) by a dedicated person: the Release Manager [RM]. PR authors and shepherds can signal that a PR should be backported by using an appropriate milestone, but ultimately, the decision is made by the RM.

The first author was the first RM in charge of putting in place this backporting process, and decided to rely on a GitHub Project board to track the status of PRs (backport requested / backported). In 2018, after creating the bot, and while he was still the RM in charge, he decided to automate the PR tracking, so that the RM could focus on taking backporting decisions and performing them.

\subsubsection{Feature}

The bot uses the milestone of the merged PRs and some information that is stored in the description of this milestone to detect PRs for which backporting was required, and it adds them to the appropriate board column. When a PR is backported to the release branch, it moves the PR to the appropriate column to indicate this updated status (see Figure \ref{fig:merge-backport}).

\subsubsection{Feedback and evolution}

Since then, the RM has been a rotating position in the Coq team. Four other developers have been in charge of backporting PRs to a release branch and they have all used this process and the automated GitHub Project board, with little change to the process or the automation.

To adapt to a RM who was particularly strict with regards to what he accepted to backport, the automation was extended to also handle backport rejections. When a PR is removed from the board by the RM, this means that they decided not to backport it. In this case, the bot changes the milestone of the PR and posts a comment to inform PR stakeholders of the decision.

\subsection{Triggering a bug minimizer}

\subsubsection{Context}

A tool to minimize test cases leading to bugs (or unexpected behavior) has been available for many years for Coq~\cite{gross_bug_min}. When they discover bugs, users sometimes make use of this tool to get to a reasonably sized test case that they can include in a GitHub issue.

\subsubsection{Feature}

To help contributors reporting issues get minimal reproducible test cases, we have integrated the bug minimizer in the Coq bot. This allows anyone to request minimization by posting a comment (or a new issue) with some script to reproduce the initial bug (e.g., downloading and building a Coq project from a git repository). When triggered, the bot will run the minimizer and post the obtained reduced test case in a comment in reply to the triggerer.

\subsubsection{Feedback and evolution}

While this feature has been documented and used, it is hard to discover for new users and also hard to use (because of the need to write a script to download a project leading to a bug).
We have extended the feature to also be able to minimize compatibility issues detected on reverse dependencies tested in Coq's CI. In this case, the bot automatically identifies the potential for minimization and proposes to trigger it. This makes the feature much more discoverable, and it has already been used a lot, despite being only a few months old.
We are now in the process of evaluating it and adapting it to the feedback that we have already received.

\section{MAINTAINING AND EVOLVING A BOT}

Our bot is project-specific: it was built to assist the Coq team, and to evolve based on their feedback. To facilitate its evolution and the involvement of any Coq developer in the bot maintenance, we chose to rely on familiar technology, and to design an easy to understand and easy to extend architecture.

\subsection{Familiar technology}

A standard choice to develop GitHub bots is Probot~\cite{probot}, a Node.js framework for GitHub Apps. However, we decided to write the bot using OCaml~\cite{leroy:hal-00930213} instead.
Indeed, this is the programming language used to build Coq, which means that Coq developers are already familiar with it. This is also a strongly-typed language, thus it provides high confidence when introducing and refactoring code, which is something else that Coq developers are quite used to.

To maximize productivity, the bot depends on many OCaml libraries (to set up a web server, encode and decode JSON, etc.). This is standard software engineering practice, but this contrasts with the practice followed in the Coq codebase, where any new dependency is carefully evaluated to guarantee stability and facilitate distribution.

Among the dependencies, graphql-ppx~\cite{graphql-ppx} is used to interface with GitHub's GraphQL API~\cite{github_graphql_API}. This API enables querying for exactly the information we need, while reducing the number of requests, and providing more safety on the request correctness thanks to the typed GraphQL language and API~\cite{graphql_spec}.

\subsection{Straightforward and extensible architecture}

The bot is architectured around a library of reusable bot components, and an application of bot workflows. The bot components are reusable bricks that can be combined into different workflows, following trigger-action programming.

Trigger-action programming~\cite{huang2015supporting} is a programming model that has mostly been studied in the context of smart-home automation, with the idea of providing a programming framework and mental model that is accessible to anyone.
Famous trigger-action programming platforms are IFTTT and Zapier~\cite{rahmati2017ifttt}.
Interestingly, both provide GitHub integration (Zapier also provides GitLab integration), but their triggers and actions are not sufficiently advanced for our purposes.

Nowadays, another example of trigger-action programming is GitHub workflows~\cite{github_actions}, which are built by combining event triggers with prebuilt or custom ``GitHub Actions''. Many tasks bots perform can also be programmed using GitHub workflows~\cite{kinsman_how_2021} but with lower reactivity, because workflows need to start virtual machines to react to events.

Our bot components are divided into the three usual types of trigger-action programming~\cite{huang2015supporting}:
\begin{itemize}
\item \emph{Event triggers}: events that the bot listens to, by subscribing to GitHub / GitLab webhooks. For instance, a new comment is an event trigger that is reused in several workflows.
\item \emph{State triggers}: additional data needed to perform an action, obtained by querying web APIs. It does not need to exactly match a function from the API. For instance, a test that a user belongs to a given team is a state trigger.
\item \emph{Actions}: state-changing requests that are sent by the bot, acting as an agent on the platform. For instance, adding a label on an issue or PR is an action that is reused in several workflows.
\end{itemize}

Introducing new bot workflows is as easy as combining the various available components, or introducing new ones when needed. The use of GraphQL to interact with GitHub makes it easy to add state triggers or actions safely.

\subsection{Team involvement}

This architecture, the use of a language that the Coq team already masters, and of external libraries as often as needed, have helped the onboarding of new bot maintainers: the first author is the initial developer and maintainer of the Coq bot since 2018; the second author was an undergraduate summer intern in 2020 who significantly extended the bot with new features and helped complete the envisioned architecture; the three other authors are Coq developers who improved and extended the bot, sometimes with little help from the initial developer.

\section{CONCLUSION}

While more and more projects adopt off-the-shelf bots to help them automate their everyday tasks, our experience adopting and maintaining a project-specific bot shows that this approach can be a successful alternative to boost developers while avoiding disruption of their pre-established workflows. Familiar technology and straightforward and extensible architecture choices can ease the maintenance of a custom bot, enabling everyone in the team to participate in its evolution.

As of today, our library of bot components is not used beyond the Coq bot application, but we could consider publishing it and extending it to fit the needs of other teams willing to write a project-specific bot in OCaml, following our example.



\section{ACKNOWLEDGMENT}

We thank Jean-Rémy Falleri and Thomas Degueule for providing feedback on early drafts of this paper.

\bibliographystyle{ieeetr}
\bibliography{biblio}

\end{document}